\documentstyle[twocolumn,aps]{revtex}
\begin{document}

\preprint{\today }
%
%
\title{Magnetic Field Effects on the Far-Infrared Absorption in Mn$_{12}$-acetate}

\author{A. B. Sushkov$^*$, B. R. Jones and J. L. Musfeldt$^*$}
\address{Department of Chemistry,  
State University of New York at Binghamton \\ 
Binghamton, New York 13902--6016}

\author{Y. J. Wang}
\address{National High Magnetic Field Laboratory,
Florida State University \\
Tallahassee, Florida 32306}

\author{R. M. Achey and N. S. Dalal}
\address{Department of Chemistry, Florida State University\\ Tallahassee, Florida 32310}
\maketitle

\begin{abstract}

We report the far-infrared spectra
of the molecular nanomagnet
Mn$_{12}$-acetate (Mn$_{12}$) as a function of temperature (5--300~K) and
magnetic field (0--17~T).
The large number of observed vibrational
modes is related to the low symmetry of the molecule,
and they are grouped together in clusters.
Analysis of the mode character based on molecular dynamics
simulations and model compound studies
shows that all vibrations are
complex; motion from a majority of atoms in the molecule contribute to most
modes.
Three features involving intramolecular vibrations of the Mn$_{12}$
molecule centered at
284, 306 and 409~cm$^{-1}$ show changes with applied magnetic field. 
The structure near
284~cm$^{-1}$ displays the largest deviation with field and is
mainly intensity related.
A comparison between the temperature dependent absorption difference
spectra, 
the gradual low-temperature cluster framework distortion as assessed
by neutron diffraction data, 
and field dependent
absorption difference spectra suggests
that this mode may involve Mn motion
in the crown. 
\end{abstract}
\narrowtext

\newpage

\section{Introduction}

Molecular magnet materials have attracted 
a great deal of interest in recent years, 
exhibiting fascinating properties such as 
cooperative phenomena, magnetic memory,
quantum tunneling, and unusual relaxation behavior that are 
most commonly associated with mesoscopic
solids. One prototype single molecule magnet is 
[Mn$_{12}$O$_{12}$(CH$_3$COO)$_{16}$\-(H$_2$O)$_4$]$\cdot$2CH$_3$COOH$\cdot$
4H$_2$O,
denoted Mn$_{12}$.
It consists of eight Mn$^{3+}$ ($S=2$)
and four Mn$^{4+}$ ($S=3/2$) ions,
held together by oxygen atoms,
acetate ligands, and waters of crystallization; the ferrimagnetic spin arrangement
yields $S$=10 \cite{Christou}.
Mn$_{12}$ crystallizes in a tetragonal lattice, with
weak exchange coupling and no long range magnetic ordering
\cite{Lis,Robinson1,Robinson2}.
Though a number of efforts have been made to understand
the energetics of the Mn$_{12}$ system, a quantitative theory 
is still lacking.
Recent prospective Hamiltonians include 
anisotropy, Zeeman splitting, spin-phonon interaction, 
and transverse terms, as well as
spin operators up to fourth order \cite{Teemu,Loss,Politi,fort}.

Mn$_{12}$ initially attracted attention due to the striking
steps and hysteresis loop in the magnetization, 
indicative of quantum tunneling.
At present, these steps are thought to result from the double 
well potential that separates spin states;
magnetization reorientation transitions have optimal 
probability when
the ``spin up'' and ``spin down'' levels of
the different magnetic quantum numbers 
align with applied magnetic field
\cite{Friedman,Barbara,Novak,Thomas2,Fried1,Fried2,Barco}.
High field EPR\cite{Hill,Barra}, neutron scattering\cite{Mire}, 
and sub-mm \cite{Mukhin} techniques
have been used to measure the excitation energies between levels in
these magnetic clusters. 
The highest energy excitation (m$_s$=10 $\rightarrow$ m$_s$=9)
occurs in the very far infrared, near 10~cm$^{-1}$ (300 GHz). 
That the magnetic dipole transition
energies in the Mn$_{12}$ system are irregular (especially near
the top of the anisotropy barrier) clearly shows the presence of
higher than second order terms in the spin Hamiltonian\cite{Hill,Mire}. 
Early heat capacity measurements revealed the irreversible/reversible effects
between the two wells below/above the blocking temperature\cite{Fomi}. 
The exact value of the blocking temperature, 
$T_b$ ($\approx$ 3~K), depends on the
probe, due to the dynamic nature of the blocking process. 
Above 3~K, the magnetization relaxation is exponential in time
and reversible; this is the thermally activated regime.
Notable deviations from exponential relaxation are observed
below 2~K \cite{Evang}.

Despite an explosion of interest in the low-energy quantum behavior,
little is known about the vibrational
characteristics of Mn$_{12}$ and many other prototypical molecular magnet materials.
Spin-phonon coupled modes have been investigated in other transition metal
oxides such as EuO, LaMnO$_3$, $\alpha$$^\prime$-NaV$_2$O$_5$, and CuO; 
the frequencies at which they appear
correspond to vibrational modulation
of the superexchange integral in the materials 
\cite{Balten,Granado,Danilo,Sherman,Kuzmenko,Kuzmenko2}, 
rather than the aforementioned
low-energy, phonon-assisted relaxation processes.
In order to provide further information on the electrodynamic
response of single molecule molecular magnets,
we have measured the far-infrared
spectra of Mn$_{12}$ as a function of both temperature
and applied magnetic field in the thermally activated regime.
We use this data to assess spin-phonon coupling in this material.  

\section{Experimental}

High quality single crystals of Mn$_{12}$ were synthesized
following the original procedure described by Lis\cite{Lis}.
The single crystals of Mn$_{12}$ were ground with paraffin at
77~K to prepare pellets of various concentrations suitable for
transmittance measurements in the far infrared \cite{Footy}. A sample with
$\approx$3.5\% of Mn$_{12}$ by mass proved to be optimal for most of the
frequency range under investigation. A more concentrated pellet ($\approx$85\%)
was used for measurements between
30 and 110~cm$^{-1}$.

Infrared transmission measurements were performed 
in our laboratory and at the National High Magnetic Field Laboratory
(NHMFL) in Tallahassee, Florida, using a Bruker 113V Fourier transform
infrared spectrometer. Spectra were taken using the 3.5, 12, 23,
and 50 $\mu$m mylar beam splitters, covering a frequency range
of 25--650~cm$^{-1}$.
Both absolute and relative
transmittance spectra were measured as a function of temperature 
using a bolometer detector and a continuous flow helium
cryostat.  Small differences in the transmittance spectra were assessed
at low temperature using absorption differences, calculated
from the relative transmittance as
$\alpha(T)-\alpha(T=5~K) = -\ln[{\cal T}(T)/{\cal T}(T=5~K)]$.
The magnetic field dependence
of the transmittance at 5~K was measured at the NHMFL using a 20~T
superconducting magnet and a transmission probe equipped with a
bolometer detector\cite{Wang,Field}.
Again, transmittance ratio spectra were
used to investigate small deviations from unity due to 
the applied field.
By taking the
natural log of the transmittance ratio
$(-\ln[{\cal T}(H)/{\cal T}(H=0)])$, we obtain the absorption
difference at each field: $\alpha(H)-\alpha(H=0)$.

Upon examination of the absorption difference curves over the entire
aforementioned frequency range, we identified three features
that change with applied magnetic field. In order to
distinguish signal from noise in a quantitative way, we
calculated the standard deviation from
the mean for each feature.
In addition, the same analysis was performed on
representative nearby frequency ranges containing no field-dependent features.
We use these standard deviations from the mean to
quantify the effects of the field and to characterize the
intrinsic noise level in regimes away from the magneto-optic signatures \cite{Stand}.
Field sweeps on an empty hole were
also carried out for reference; as expected, no
field dependence was observed.

\section{Results and Discussion}

\subsection{Temperature Dependence of the Far Infrared Vibrational Spectra}

Figure 1 displays the far infrared transmittance of Mn$_{12}$ as a
function of temperature. As expected for a molecular solid,
intramolecular vibrational modes of the Mn$_{12}$ molecule appear
above 150~cm$^{-1}$ \cite{Dresselhaus}.
In agreement with
previous authors \cite{Mukhin,Hennion}, we also
observe a weak low-energy structure centered near
35 cm$^{-1}$ (Fig. 1, panel a).
This magnetic-dipole allowed excitation
has been cited as evidence for other excited state multiplets
with $S$$\ne$10 existing in Mn$_{12}$ \cite{Mukhin,Hennion}.
Within our sensitivity, we have found that the 35 cm$^{-1}$ feature 
displays limited temperature dependence
and no magnetic field dependence.

The far infrared spectra of Mn$_{12}$ display a large number of intramolecular
modes due to the relatively low symmetry of the molecule. They are grouped
 together in clusters and superimposed upon one another. These structures
 sharpen and harden
with decreasing temperature (Fig. 1).
Recent 20 K 
neutron diffraction studies by Langan $et~al.$ \cite{Robinson1}
assessed the low-temperature molecular distortion; compared to the
300 K structure, major characteristics
include  a gradual displacement of Mn(3) and carboxylate groups, and more extensive
solvent interactions including a 4-center hydrogen bond. 
(Note that Mn(3) is part of the crown, according to the numbering scheme of
Ref. \cite{Robinson1}, with
connections/interactions to the bridges,
ligands, and solvent.) 
These results motivated us to more
carefully investigate the far infrared
response in the low-temperature regime.

Figure \ref{temp}a
displays the absorption difference
spectra of Mn$_{12}$ at low temperature \cite{noteT}. 
The gradual distortion of the cluster framework was previously
assessed by neutron diffraction 
studies \cite{Robinson1}. From a magnetic properties point of
view, the behavior of the Mn ions is most important;
experiment indicates that Mn(3) on the crown displays the most
significant low-temperature displacement \cite{Robinson1}.  
A number of vibrational
features  
are modified in this temperature range 
and therefore likely contain a substantial Mn(3),
carboxylate, and acetate ligand contribution.
As will be discussed in the next Section,
three modes (284, 306, and 409~cm$^{-1}$) are of particular interest because of
their dependence on magnetic field.
The 284~cm$^{-1}$ mode shows 
an intensity variation at low temperature 
and a small frequency shift at higher
temperatures (Fig. \ref{temp}b).
The standard
deviation from the mean in the vicinity of the 284~cm$^{-1}$ 
feature (Fig. \ref{temp}c) 
quantifies these changes, with an inflection point
around 20~K. 
The maximum in the derivative of
the fourth-order polynomial fit to this data better illustrates the
position of this inflection point.
The 306~cm$^{-1}$ structure (not shown) is sensitive to this
temperature regime as well.
Although the feature centered at 409~cm$^{-1}$ also exhibits
changes in this temperature range, no peculiarities
were observed around 20~K.

Preliminary molecular dynamics simulations \cite{Simulation}
suggest that
a majority of atoms in the Mn$_{12}$ molecule are, in
some way, involved in each vibration in this frequency range.
The modes are many, both because
of the low molecular symmetry and the large number of atoms.
Using mode visualization
and the spectra of several model compounds as a guide, particularly that of
Mn(II)-Ac$_2$ (Fig. \ref{model}), we
propose the following general assignments.
The low-energy motions below 300~cm$^{-1}$
include a great
deal of acetate (ligand) motion. Complex low energy
motions of the core and
crown begin around 180~cm$^{-1}$. These motions include bending, rocking,
shearing, twisting,
and wagging. Asymmetric and symmetric
stretching of the core and crown seem to begin slightly below 500~cm$^{-1}$,
thus providing a likely
assignment for the complicated vibrational cluster observed in the
spectrum centered near 540~cm$^{-1}$.
The simulations show that modes in this energy range
contain sizable (core and crown) 
oxygen contributions. Comparison of the spectral data from the model
compound Mn(II)-Ac$_2$ and Mn$_{12}$ 
confirm that the mode clusters centered near 360, 400, and 540~cm$^{-1}$ in
Mn$_{12}$ are mainly motions of the magnetic center rather than the ligands.
Related model compounds such as MnO, MnO$_2$, and Mn$_2$O$_3$ also have
characteristic vibrational features in the far infrared (below 600~cm$^{-1}$).
Although the structures are different and Mn valences unmixed compared to the
title compound, the reference spectra of these model
Mn-based solids \cite{Stadtler} indicate that the
far infrared response is rich and highly relevant to the Mn--O motion.
Measurements on the Mn(II)-Ac$_2$ model compound (Fig. 3) suggest that the
structures centered at 590 and 635~cm$^{-1}$  in the
spectrum of Mn$_{12}$ are acetate-related. The similar temperature dependence
of these modes in the two samples supports this assignment.

In order to understand and model physical properties that contain
important phonon contributions (for instance $\rho$$_{DC}$
or heat capacity) in the thermally
activated regime, it is helpful to know realistic
values of the most
important low-energy infrared active phonons of Mn$_{12}$. 
Our data suggests that use of
a relatively small series
of characteristic frequencies
may be adequate. For instance, the three most intense structures are centered at
375, 410, and 540~cm$^{-1}$.
More complete models might
take additional modes, the detailed
mode clustering, and relative intensities into account.

\subsection{Magnetic Field Dependence of the Far
Infrared Vibrational Spectra}

Figure 4 displays the magnetic field dependence of three features in
the far infrared spectrum of Mn$_{12}$. The center positions of these structures
($\approx$284, 306, and 409~cm$^{-1}$) are
indicated with arrows on
the absolute transmittance spectra in Fig. 1. It is interesting that the
284 and 306~cm$^{-1}$ features do
not correspond to major modes in the absolute transmittance spectrum.
As shown in the left-hand
panels (Figs. 4a, 4c, and 4e), the
field dependence as measured by the absorption difference spectra is well-defined for
the mode near 284~cm$^{-1}$,
whereas it is more modest for the structures centered at
306 and 409~cm$^{-1}$.  
The filled symbols and solid lines in the right-hand panels
(Figs. 4b, 4d, and 4f) provide a
quantitative view of the field dependent trends. 
The standard deviation from the mean for  
the three modes of interest
shows a clear upward trend, in contrast to
that of nearby spectral regions 
which shows no field
dependence and provides an estimate of the intrinsic noise level
\cite{Deviation}.

The feature at 284~cm$^{-1}$ is the most prominent
of the three field dependent structures.
The absorption decreases with applied field
and shows a tendency toward saturation at 17~T, as indicated in 
Fig. \ref{ratio}b.
The standard deviation from the mean in nearby spectral ranges
(for instance, similarly sized regimes centered at $\approx$ 277 
and 299~cm$^{-1}$) shows no
field dependence.
The response of the 284~cm$^{-1}$ feature
is therefore well above the noise and characteristic of Mn$_{12}$.
Similar trends and a tendency towards saturation
of the standard deviation from the mean are observed for
the more complicated absorption difference structures
centered near 306 and 409~cm$^{-1}$, although
the overall coupling of these modes to the applied field is weaker and the
line shapes are
different from that of the 284~cm$^{-1}$ feature.
The changes observed near 306~cm$^{-1}$ are mainly caused by a 1~cm$^{-1}$
softening, whereas the 409~cm$^{-1}$ structure
is due to both softening and broadening.
The field dependence of these three features in the absorption difference
spectra is confirmed by repeated measurements using different
experimental parameters \cite{Parameters}.
Further, as seen in Fig.~4,
the trend with applied field in both the difference spectra and 
the standard deviation from the mean
near 306 and 409~cm$^{-1}$ can be easily distinguished
from that of the non-field-dependent intervals.

It is of interest to understand the microscopic
character of the 284, 306, and 409~cm$^{-1}$
modes and why they are affected by the magnetic field.
Although there are a number of mechanisms that might explain this
behavior, such as the effect of magnetic field on hybridization of
ligand-metal bonds \cite{Brunel}, we believe the most promising is that of
spin-phonon coupling.
Here, variations in the applied field modulate the magnetic system,
thereby affecting the phonons coupled to it.
Such an interaction has been thought to be important for
describing the slow magnetization relaxation behavior of Mn$_{12}$, making
it an important term in the total
Hamiltonian \cite{Politi,fort}.  That only three far-infrared features
are observed to change with applied field up to 17~T suggests that these
modes are related to
motions that maximally change the low temperature
spin interactions.
One possible candidate
for such motion might contain bending between the core and crown
structures affecting the Mn-O-Mn angle, therefore modulating the
superexchange between Mn ions.
An alternate explanation involves the especially curious 
sensitivity of the 284 cm$^{-1}$ mode to temperature variation (in the range 5-35 K)
as well as to the applied magnetic field.
This coincidence suggests that the magneto-elastic response at 284~cm$^{-1}$
may be intimately related to the Mn(3) motion on the crown. Interestingly,
the 306 cm$^{-1}$ structure also displays some sensitivity to the
displacement of the molecular framework. 
Further work is clearly needed
to untangle these complex interactions.

The observed magnetic field dependencies at 284,
306, and 409~cm$^{-1}$ suggest that ${\mathcal H}_{s-ph}$ can be augmented by
including contributions from these three vibrational
modes. Spin-lattice interactions
in the far-infrared energy range have been observed in a number of other transition
metal oxides in the past \cite{Balten,Granado,Danilo,Sherman,Kuzmenko,Kuzmenko2}. However, the energy
scale of the 284, 306, and 409 cm$^{-1}$ features is
larger than might be expected for relevant
spin-phonon processes in Mn$_{12}$, even in the thermally activated regime.
Indeed, previous studies on the title compound have focused
on acoustic phonons as the major contributors to
${\mathcal H}_{s-ph}$ \cite{Teemu}. Thus, if these
far-infrared phonons are connected with the Mn$_{12}$
relaxation processes in any
way, virtual states would likely be involved. 
The energy scale of the 284, 306, and 409~cm$^{-1}$ phonons may be more
relevant to high field transport or specific heat measurements.

\section{Conclusion}

We have measured the far-infrared response of Mn$_{12}$ as a function of
temperature and magnetic field.
The features in this region involve vibrations
from all of the groups of atoms (core, crown, and ligands) in the molecule.
Three structures related to intramolecular vibrations
near 284, 306, and 409~cm$^{-1}$
are observed to change in applied magnetic field,
suggesting that they are coupled to the spin system.
Of these
three, the feature centered at 284~cm$^{-1}$ displays
the strongest coupling. 
Based on the similarity between the temperature and field dependent
absorption difference data, we speculate that this mode involves Mn(3) motion
on the crown. 
The data reported here may be helpful in understanding the role of
vibrations in the theoretical models of magnetization relaxation in
Mn$_{12}$ and related systems.

\vspace{0.4in}
\noindent
$^*$ Current address: Department of Chemistry, University of Tennessee,
Knoxville, TN 37996.

\vspace{0.4in}

\section{Acknowledgements}

Funding from the National Science Foundation (DMR-9623221) to support
work at SUNY-Binghamton is gratefully acknowledged.
The work at Florida State University was also partially
funded by the NSF.
A portion of the measurements were performed at the NHMFL,
which is supported by NSF Cooperative Agreement No. DMR-9527035
and by the State of Florida. We thank Z.T. Zhu for
technical assistance.



\begin{figure}
\caption{Far infrared transmittance spectra 
of 85\% by mass (panel a) and 3.5\% by mass (panels b, c, and d)
Mn$_{12}$
suspended in a paraffin pellet
at various temperatures (10, 100, 200 and 300~K).
Arrows indicate the three magnetic field dependent features at 284,
306, and 409~cm$^{-1}$. 
\label{fir}}
\end{figure}

\begin{figure}
\caption{ a) Far-infrared absorbance difference spectra
($\alpha(T)-\alpha(T=5~K)$) of Mn$_{12}$ as
a function of frequency at low temperature. The curves
are offset for clarity. The arrow indicates 10\% deviation from unity.
b) Close-up view of the absorbance difference spectra
near the 284~cm$^{-1}$ feature at low temperature. Dashed lines: temperatures
below 20~K; solid lines: temperatures above 20~K.  c) Standard deviation from the
mean of the 284~cm$^{-1}$ absorbance difference.  Solid line: fourth-order
polynomial fit to guide the eye; dashed line: derivative of fourth-order
polynomial fit.
\label{temp}}
\end{figure}

\begin{figure}
\caption{Far-infrared transmittance spectra of Mn$_{12}$ alongside
similar data on the model compound Mn(II)-Ac$_2$.  Spectra were collected at
10~K.
\label{model}}
\end{figure}

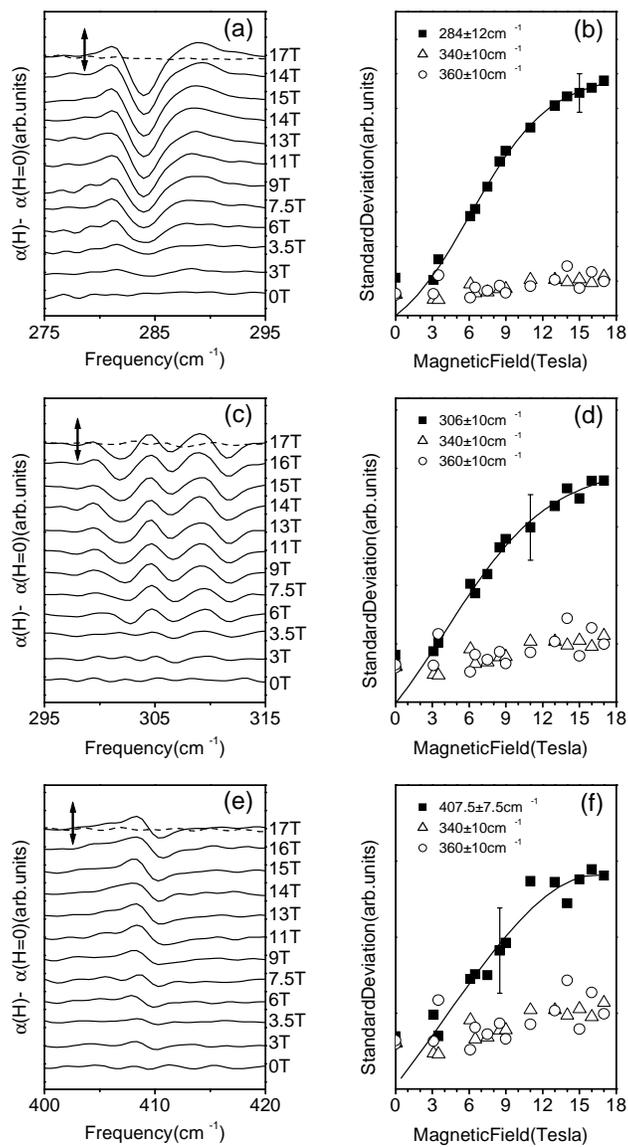
\begin{figure}
\caption{Magnetic field dependent features in absorbance difference
spectra at 5~K.
Left column of plots: $\alpha(H)-\alpha(H=0)$ vs. frequency at different
magnetic fields (curves are shifted for clarity) for features
at 284 (a), 306 (c), and 409~cm$^{-1}$ (e). Solid curves: sample;
dashed curve: empty reference (hole).  Arrows indicate 5\% deviation from unity.
Right column: standard deviation from the mean value vs. magnetic field
calculated for the frequency intervals indicated on the plots (b, d, and f).
Solid lines guide the eye.
\label{ratio}}
\end{figure}


\begin{references}
\bibitem{Footy} Note that this process results in an isotropic sample.
\bibitem{Field} The field dependence was measured in the following order:
0, 9, 13, 15, 17, 16, 14, 11, 8.5, 6.5, 3.5, 0, 3, 6, 7.5, and 0~T.
No hysteresis was observed.
\bibitem{Stand} Standard deviation from the mean was used to quantify the features
in the absorbance difference spectra rather than integrated intensity due to the
complicated line shapes of some of the features.  This method was also used to
estimate the intrinsic noise level by analysis of nearby spectral regions.
Thus, for consistency, standard deviation from the mean was used for all
treatments of the difference spectra.
\bibitem{noteT} This temperature range is relevant to the energy
scale of the magnetic measurements.
\bibitem{Simulation} Molecular dynamics simulations were performed using
the MMFF method of the Titan program by Schrodinger, Inc..
\bibitem{Deviation} The method for determining the standard deviation from the mean
is described in the Experimental section.
\bibitem{Parameters} The field dependences of these three features 
were confirmed by repeating the measurements using different
(yet valid) scan velocities and beamsplitters.
\end{references}
\end{document}